\documentclass[a4paper,fleqn,usenatbib]{mnras}
\usepackage[T1]{fontenc}
\usepackage{ae,aecompl}
\usepackage{times}

\usepackage{graphicx}	
\usepackage{amsmath}	
\usepackage{amssymb}	
\usepackage{color}
\usepackage{mathptmx}
\usepackage{epsfig}
\usepackage{wasysym}
\usepackage{xfrac}
\usepackage{pifont}

\title[Are mergers important for stellar mass growth?]{The limited role of galaxy mergers in driving stellar mass growth over cosmic time}

\author[G. Martin et al.]{
G. Martin,$^{1}$\thanks{E-mail: g.martin4@herts.ac.uk}
S. Kaviraj,$^{1}$
J. E. G. Devriendt,$^{2}$
Y. Dubois,$^{3}$
C. Laigle$^{2}$
and C. Pichon,$^{3,4}$ 
\\
$^{1}$Centre for Astrophysics Research, School of Physics, Astronomy and Mathematics, University of Hertfordshire, College Lane, Hatfield AL10 9AB, UK\\
$^{2}$Dept of Physics, University of Oxford, Keble Road, Oxford OX1 3RH UK\\
$^{3}$Institut d'Astrophysique de Paris, Sorbonne Universit\'{e}s, UMPC Univ Paris 06 et CNRS, UMP 7095, 98 bis bd Arago, 75014 Paris, France\\
$^{4}$Korea Institute of Advanced Studies (KIAS) 85 Hoegiro, Dongdaemun-gu, Seoul, 02455, Republic of Korea
}

\pubyear{2017}

\begin{document}
\label{firstpage}
\pagerange{\pageref{firstpage}--\pageref{lastpage}}
\maketitle

\newcommand{\fmajor}{10 }
\newcommand{\fminor}{15 }
\newcommand{\fall}{25 }
\newcommand{\frest}{75 }

\begin{abstract}
A key unresolved question is the role that galaxy mergers play in driving stellar mass growth over cosmic time. Recent observational work hints at the possibility that the overall contribution of `major' mergers (mass ratios $\gtrsim 1:4$) to cosmic stellar mass growth may be small, because they enhance star formation rates by relatively small amounts at high redshift, when much of today's stellar mass was assembled. However, the heterogeneity and relatively small size of today's datasets, coupled with the difficulty in identifying genuine mergers, makes it challenging to \textit{empirically} quantify the merger contribution to stellar mass growth. Here, we use Horizon-AGN, a cosmological hydrodynamical simulation, to comprehensively quantify the contribution of mergers to the star formation budget over the lifetime of the Universe. We show that: (1) both major and minor mergers enhance star formation to similar amounts, (2) the fraction of star formation directly attributable to merging is small at all redshifts (e.g. $\sim$35 and $\sim$20 per cent at z$\sim$3 and z$\sim$1 respectively) and (3) only $\sim$\fall per cent of today's stellar mass is directly attributable to galaxy mergers over cosmic time. Our results suggest that smooth accretion, not merging, is the dominant driver of stellar mass growth over the lifetime of the Universe. 
\end{abstract}

\begin{keywords}
methods: numerical -- galaxies: evolution -- galaxies: formation -- galaxies: high-redshift -- galaxies: interactions
\end{keywords}

\section{Introduction}
Understanding the processes that drive stellar mass growth over cosmic time is a key topic in observational cosmology. Since the cosmic star formation rate (SFR) density peaked at $z\sim2$ and dropped by more than an order of magnitude towards the present day \citep[e.g.][]{Madau2014,Gonzalez2014}, almost half of the stellar mass hosted by today's galaxies formed at $z \gtrsim 1.3$ \citep{Madau2014}, making this epoch particularly important in the evolution of the observable Universe.

Galaxy mergers are often considered to be important drivers of stellar mass growth \citep[e.g.][]{Dokkum2010,Kaviraj2011,Lopez2012,Ferreras2014}. For example, mergers can produce orders-of-magnitude enhancements in SFRs in the nearby Universe \citep[e.g.][]{Duc1997,Elbaz2003}, implying that a significant fraction of the stellar mass formed in these episodes is a direct consequence of the merger event. Since the merger rate increases towards high redshift, it is reasonable to consider whether a significant fraction of the stellar mass in today's galaxies may, therefore, have been created in enhanced star-formation episodes associated with galaxy mergers \citep[e.g.][]{Somerville2001,Conselice2008}. In other words, if galaxy mergers are frequent and routinely enhance SFRs when they take place, then much of the stellar mass at the present day could be directly attributable to the merging process.

However, while mergers are clearly capable of triggering bursts of star formation \citep[e.g.][]{Mihos1996,DiMatteo2008}, and strongly star-forming systems are often coincident with  ongoing interactions \citep[e.g.][]{Sanders1988,Bell2006}, the empirical picture remains unclear, especially at high redshift. Recent observational studies of galaxies around the epoch of peak cosmic star formation \citep[e.g.][]{Rodighiero2011,Stott2013,Lofthouse2017} indicate that `major' mergers (i.e. mergers with mass ratios $>1:4$) are unlikely to be responsible for the bulk of the stellar mass growth at these epochs, as the SFR enhancements in major mergers -- compared to the non-merging population -- are relatively low \citep[e.g.][]{Lofthouse2017}. This implies that there must be other processes that fuel these high SFRs and drive the production of stellar mass at these epochs. 

Given that the frequency of `minor' mergers (mass ratios $<1:4$) is several times that of major mergers \citep[e.g.][]{Lotz2011,Kaviraj2015}, and that mergers of moderate mass ratios are also capable of producing large SFR enhancements \citep[e.g][]{Cox2008}, minor merging could potentially make an important contribution to the star formation budget \citep{Kaviraj2014a,Kaviraj2014b}. Alternatively, the high SFRs may simply be the result of high molecular-gas fractions, fuelled by intense cosmological gas accretion \citep[e.g.][]{Tacconi2010,Geach2011,Bethermin2015}.

While quantifying the role of mergers in driving cosmic stellar mass growth is an important exercise, an empirical determination of this issue brings with it several difficulties. Selecting mergers based on morphological disturbances is not a simple task, since disturbed morphologies can also result from internal processes, especially in the early Universe \citep[e.g.][]{Bournaud2008,Agertz2009,Forster2011,Cibinel2015,Hoyos2016}. Furthermore, since the surface brightness of merger-induced tidal features declines with the mass ratio of the merger, minor mergers are less likely to produce observable asymmetries, especially at high redshift, even in today's deep surveys \citep{Kaviraj2013}. Finally, given the depth and areal coverage of current and past facilities, samples of mergers are often small, and both the galaxy populations studied and star formation indicators employed can be heterogeneous, making it difficult to compare results across a large range in redshift. 

With these issues in mind, an appealing alternative is to employ a simulation that reproduces the observed properties of galaxies over cosmic time \citep[e.g.][]{Lamastra2013,Vogelsberger2014,Schaye2015,Khandai2015,Taylor2016,Kaviraj2017}. A major advantage of this approach is that, since the identities of the progenitors of each galaxy in the simulation are precisely known, it is straightforward to separate merging galaxies from their non-merging counterparts. This then allows us to integrate over the star formation history of each merger (including any subsequent post-starburst decrease in the SFR), study the properties of the induced star formation and make quantitative statements about the overall role of merging in creating the stellar mass in today's Universe.

In this Letter, we use the hydrodynamical cosmological simulation, Horizon-AGN\footnote{\url{http://www.horizon-simulation.org}} \citep{Dubois2014,Kaviraj2017}, to quantify the contribution of mergers to the star formation budget since $z=6$. In Section \ref{sec:simulation}, we describe the simulation and the prediction of observable quantities in the model. In Section \ref{sec:enhancement}, we describe our identification of mergers and calculate the merger contribution to the star formation budget as a function of redshift. In Section \ref{sec:contribution}, we quantify the contribution of both major and minor mergers to the cosmic star formation history. We summarize our findings in Section \ref{sec:summary}. 

\begin{figure}
	\centering
    \includegraphics[width=0.50\textwidth]{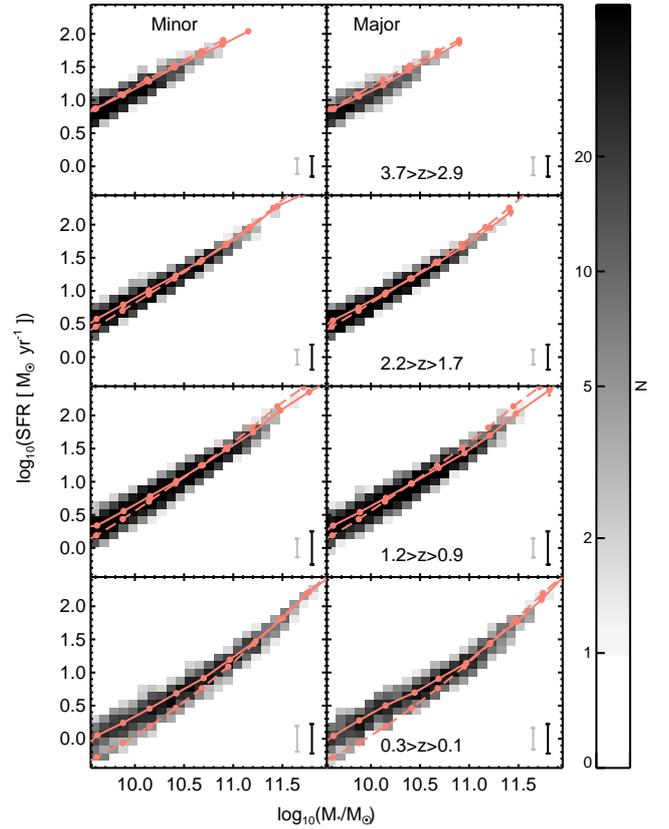}
    \caption{SFR as a function of stellar mass for the merging and non-merging populations in various redshift ranges. Greyscale density maps represent the minor (left) and major (right) merger populations. Solid pink lines shows the mean SFR of the merging population, while the dashed pink lines show the mean SFR of the non-merging population in each redshift bin. The offset between the solid and dashed lines in every panel therefore indicates the average enhancement due to a merger. The grey and black error bars indicate the typical standard deviations for the merging and non merging populations respectively.}
    \label{fig:mainsequence}
\end{figure}

\section{The Horizon-AGN simulation}
\label{sec:simulation}
We begin with a brief description of the Horizon-AGN simulation and the prediction of observable quantities in the model. Horizon-AGN is a cosmological hydrodynamical simulation \citep{Dubois2014} that employs RAMSES \citep{Teyssier2002}, an adaptive mesh refinement Eulerian hydrodynamics code. It simulates a volume of (100~$h^{-1}\rm coMpc$)$^{3}$ containing 1024$^{3}$ DM particles (M$_{\mathrm{DM}}=8\times10^{7}$M$_{\odot}$) and uses initial conditions from a \emph{WMAP7} $\Lambda$CDM cosmology \citep{Komatsu2011}. The initial gas mass resolution is $10^{7}$M$_{\odot}$, with a maximum grid refinement of $\Delta x = 1$~kpc. Horizon-AGN includes sub-grid prescriptions for star formation and stellar/AGN feedback. Star formation proceeds with a standard 2 per cent efficiency per free-fall time \citep{Kennicutt1998}, once the Hydrogen gas density reaches $n_{0}=0.1$~H~cm$^{-3}$. Continuous stellar feedback is employed which includes momentum, mechanical energy and metals from Type II SNe, stellar winds, and Type Ia SNe \citep{Kaviraj2017}, with the Type Ia SNe implemented following \citet{Matteucci1986}, assuming a binary fraction of 5 per cent. Black-hole feedback on ambient gas operates via a combination of two channels and depends on the ratio of the gas accretion rate to the Eddington luminosity, $\chi=\dot{\mathrm{M}}_{\mathrm{BH}}/\dot{\mathrm{M}}_{\mathrm{Edd}}$. For Eddington ratios greater than 0.01 (high accretion rates) a `quasar' mode is active with 1.5 per cent of the accretion energy being injected isotropically into the gas as thermal energy. For Eddington ratios less than 0.01 (low accretion rates) a `radio' mode is active, where cylindrical bipolar outflows are employed with a jet velocity of $10^4$~km~s$^{-1}$. The efficiency of the radio mode is higher, at 10 per cent of the accretion energy. The quasar mode efficiency is chosen to reproduce observed relations between M$_{\rm BH}$ -- M$_\star$ and M$_{\rm BH}$ -- $\sigma_\star$ relations as well as the local cosmic black-hole mass density \citep{Dubois2012}.

Horizon-AGN reproduces key observables that trace the aggregate cosmic stellar mass growth of galaxies: stellar mass and luminosity functions, rest-frame UV-optical-near infrared colours, the star formation main sequence and the cosmic star formation history \citep{Kaviraj2017}. It also reproduces galaxy merger histories \citep{Kaviraj2015} and the demographics of black holes (BHs): the BH luminosity and mass functions, the BH mass density versus redshift, and correlations between BH and galaxy mass \citep{Volonteri2016}.

We use the \textsc{AdaptaHOP} structure finder \citep{Aubert2004,Tweed2009}
to identify galaxies in the final snapshot of the simulation ($z=0.06$), and build merger histories for each galaxy. We produce a catalogue of galaxies with $\mathrm{M}>10^{9.5}$~M$_{\astrosun}$ from $z=0.06$ to $z=6$ and calculate the stellar mass formed in each galaxy between timesteps. Since the minimum galaxy mass identified by the structure finder is M$_{\star}\approx 2\times10^{8}$~M$_{\astrosun}$, our sample is complete for mergers down to a mass ratio of at least 1:15.

\section{Star formation enhancement due to merging}
\label{sec:enhancement}
We begin our analysis by identifying mergers in the simulation and measuring their mass ratios (Section \ref{sec:mergers}). We then compare the SFRs of merging galaxies with those of the non-merging galaxy population, so as to estimate (and `subtract') the star formation that would have taken place anyway in the absence of merging (Section \ref{sec:StarFormation}). This then enables us to calculate the stellar mass growth that is directly attributable to the merger process. 

\subsection{Defining and identifying mergers}
\label{sec:mergers}

To measure the SFR of the merging system, we calculate the total stellar mass formed in a 2~Gyr window, centred around the time that the two galaxies coalesce (i.e. when both galaxies are identified as being part of the same structure). We note that the size of the window is chosen to encompass the star formation history of the system around the merger, and that the exact choice of timescale (e.g. increasing it to 3~Gyrs or even reducing it to 1~Gyr) does not alter our conclusions. 

It is also worth noting that how the mass ratio is defined can influence the minor and major merger rate and therefore the results of such an analysis \citep[e.g.][]{Rodriguez2015}. For this study, we use the mass ratio calculated when the satellite is at its maximum mass prior to coalescence -- i.e. before material begins to be transferred between the merging companions -- because this measures the `true' mass ratio of the system, before the merger process begins to alter the properties of the merging progenitors. Only mergers with mass ratios greater than $1:10$ are considered since, in agreement with previous studies \citep[e.g.][]{Cox2008}, we find that smaller mass ratio mergers have a negligible effect on the star formation rate. 

\begin{figure}
	\centering
    \includegraphics[width=0.5\textwidth]{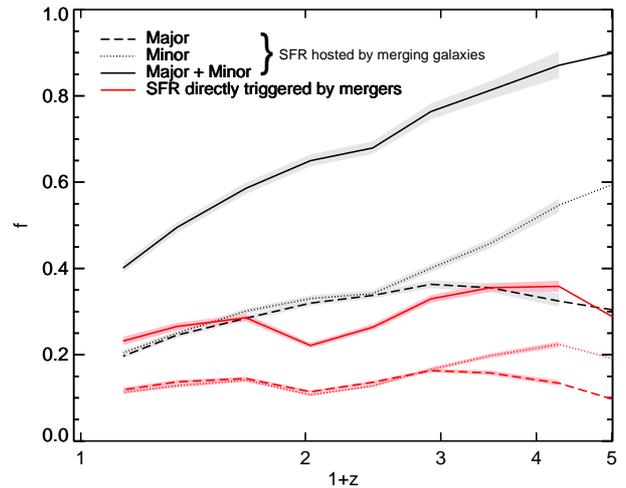}
    \caption{The fraction of star formation in merging systems (black) and the fraction of the star formation budget that is directly triggered by merging (red) as a function of redshift. Dotted lines indicate the contribution of minor mergers; dashed lines indicate the contribution of major mergers; solid lines indicate the combined contribution of major and minor mergers. Filled polygons indicate the 1$\sigma$ errors obtained from bootstrap re-sampling ($n=1000$).}
    \label{fig:budget}
\end{figure}

\begin{figure}
	\centering
    \includegraphics[width=0.5\textwidth]{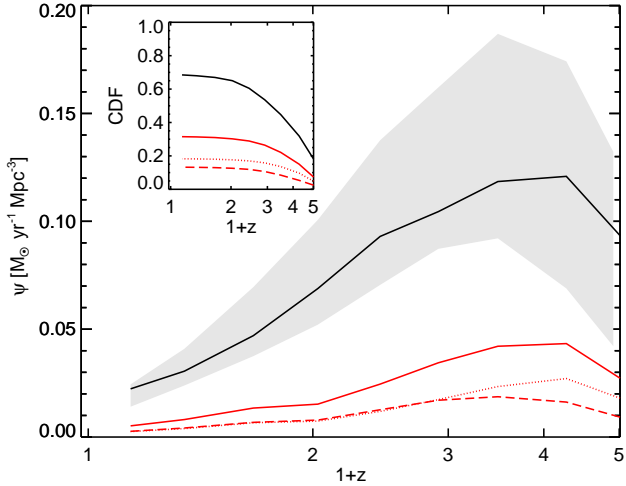}
    \caption{The cosmic star formation rate density from the Horizon-AGN simulation (black) and the contribution due to major and minor mergers (red). The inset shows a cumulative version of this plot i.e. the cumulative fraction of stellar mass formed due to mergers (red) and due to other processes (black). The grey filled area indicates the $3\sigma$ confidence region from observations \citep{Hopkins2006b}.}
    \label{fig:cosmicsfr}
\end{figure}

\subsection{Star formation triggered by mergers}
\label{sec:StarFormation}
Figure \ref{fig:mainsequence} compares the star-formation main sequence of the merging and non-merging galaxy populations in four redshift bins. The mergers are further split into minor mergers (mass ratios $< 1:4$; left-hand column) and major mergers (mass ratios $\gtrsim 1:4$; right-hand column). The solid pink lines show the mean SFRs of the merging populations, while the dashed pink lines shows the mean SFRs of the non-merging populations in each panel. The difference between the solid and dashed lines therefore indicates the (average) enhancement of star formation in merging galaxies at a given epoch. We note that, in common with other theoretical work \citep[e.g.][]{Dave2008,Lamastra2013}, the normalisation of the star-formation main sequence in Horizon-AGN is underestimated compared to its observational counterparts by $\sim0.2$ dex \citep{Kaviraj2017}. However, since the baryonic recipes used are not altered during merging, any star-formation enhancement in merging galaxies will be proportionally reduced to the same extent, leaving our conclusions unchanged.

It is interesting to note that the SFR enhancement due to minor mergers does not differ significantly from that of major mergers, consistent with the findings of recent observational studies \citep[see e.g.][]{Willett2015,Carpineti2015}. This is likely driven by the fact that the gas inflows which underpin the SFR enhancements in mergers \citep{Dimatteo2007} can be of similar magnitude in minor mergers as they are in their major counterparts \citep[e.g.][]{Hernquist1995}. Furthermore, merger-driven SFR enhancement is most efficient in the local universe, because the `ambient' level of star formation due to secular processes is much lower, which allows violent events like mergers to produce significant enhancements in the SFR \citep[e.g.][]{Mihos1996}. We define the merger-driven enhancement of star formation, $\xi$, as the ratio of the mean specific star formation rate (sSFR) in the non-merging population to that in the merging population. We measure $\xi$ in bins of both redshift and stellar mass (since the sSFR has a dependence on this parameter \citep[e.g.][]{Whitaker2012}):

\begin{equation}
	\xi(m_{*},z) = \frac{\big\langle \mathrm{sSFR}_{m}(m_{*},z) \big\rangle}{\big\langle \mathrm{sSFR}_{non}(m_{*},z) \big\rangle}.
\end{equation}

\noindent The enhancement can be used to estimate the fraction of star formation that would have occurred in the merger progenitors anyway, had they not been in the  process of merging. For example, if $\xi$ is a factor of 2 then, on average, around half the star formation in the merging system in question is likely driven by other processes \citep[see e.g.][]{Kaviraj2013b,Lofthouse2017}. By subtracting the star formation that would have occurred anyway had the merger not taken place, we can then measure the fraction of star formation that is \textit{directly due} to mergers ($f$) as follows:

\begin{equation}
	f = \frac{m_{new,m}(m_{*},z) \big[1- 1/ \xi(m_{*},z)\big]}{m_{new,total}(m_{*},z)},
\end{equation}

\noindent where $m_{new,m}(m_{*},z)$ is the total stellar mass formed in mergers in a given stellar mass and redshift bin and $m_{new,total}(m_{*},z)$ is the total stellar mass formed in the simulation in the stellar mass and redshift bin in question. 

As Figure \ref{fig:budget} shows, the fraction of star formation \textit{in} merging galaxies increases towards high redshift, reflecting the increasing merger rate. However, the fraction of star formation that is \textit{directly due} to mergers (shown by the red lines in Figure \ref{fig:budget}) does not increase to the same extent, which is a consequence of a decreasing merger-driven SFR enhancement towards high redshift, as shown in Figure \ref{fig:mainsequence}. The fraction of star formation triggered by merging peaks around $z\sim3$ ($\sim35$ per cent), and then decreases to $\sim20$ per cent by $z\sim1$. We find that, on average, 65 per cent of the enhanced star formation due to a merger takes place prior to coalescence, with the star formation rate in the post-merger remnant returning to that of the non-merging population in less than 1~Gyr for galaxies at $z>1$. It is worth noting that our results are consistent with recent observational and theoretical work that has probed the contribution of major mergers to the cosmic SFR density in selected redshift ranges. For example,  \citet{Lamastra2013} and \citet{Robaina2009} indicate that the major-merger contribution to cosmic star formation at low/intermediate redshift ($0.4<z<2$) is around 10 per cent, with only modest SFR enhancements at these epochs \citep{Robaina2009,Fensch2017}, as indicated by Figure \ref{fig:mainsequence}.

\section{The merger contribution to the cosmic star formation history}
\label{sec:contribution}
We proceed by studying the merger contribution to the overall build-up of stellar mass over cosmic time, by multiplying the fraction of star formation directly due to  mergers from Section \ref{sec:enhancement} (red lines in Figure \ref{fig:budget}) by the cosmic star formation rate density ($\psi$). We present, in Figure \ref{fig:cosmicsfr}, the cosmic star formation rate density in Horizon-AGN (black solid line). Since our sample of simulated galaxies is limited to masses above $10^{9.5}$~M$_{\astrosun}$, and the merger-driven enhancement of star formation increases for galaxies with lower stellar mass (Figure \ref{fig:mainsequence}), it is important to ask if galaxies less massive than our mass threshold could contribute significantly to the star formation budget. To explore this, we multiply the star formation rate vs stellar mass trend at $z\sim 0$ \citep{Elbaz2007} and $z\sim 2$ \citep{Daddi2007} with the galaxy stellar mass functions at the same redshifts from \citet{Baldry2008} and \citet{Tomczac2014}, in order to produce star formation rate densities per dex in stellar mass down to $10^{7}$~M$_{\astrosun}$. We find that only $\sim$22 per cent and $\sim$16 per cent of stellar mass at $z\sim 0$ and $z\sim 2$ respectively is formed in galaxies less massive than $10^{9.5}$~M$_{\astrosun}$. It appears reasonable, therefore, to assume that considering the full stellar mass range would not significantly alter our conclusions. 

Figure \ref{fig:cosmicsfr} indicates that the proportion of the cosmic star formation budget that is directly attributable to merging is small at all redshifts. Following the trends in Figure \ref{fig:budget}, it peaks around $z\sim3$ ($\sim0.04$~M$_{\astrosun}$~yr$^{-1}$~Mpc$^{-3}$) and then steadily declines towards the present day. The inset shows a cumulative version of this plot, indicating that only $\fall$ per cent of the star formation budget since $z \sim 6$ is attributable to mergers ($\sim\fmajor$ per cent from major mergers and $\sim\fminor$ per cent from minor mergers). Recall that the contribution by very low mass ratio ($<1:10$) mergers is expected to be negligible, so that this result should hold generally for all mergers over cosmic time. While a detailed study of the role of secular processes is beyond the scope of this Letter, our results indicate that an overwhelming majority ($\sim$ $\frest$ per cent) of the cosmic star formation budget is unrelated to merging and a result of secular evolution, driven simply by cosmological accretion of molecular gas, in line with the suggestions of recent observational work \citep[e.g.][]{Tacconi2010,Bethermin2015} and previous theoretical work which has suggested that non-merging systems dominate the SFR density at all redshifts \citep[e.g.][]{Hopkins2010}.

\section{Summary}
\label{sec:summary}
We have used the Horizon-AGN cosmological hydrodynamical simulation, to quantify the contribution of galaxy mergers to stellar mass growth over cosmic time. Our key results are as follows:   

\begin{itemize}
\item \textit{Mergers enhance star formation most efficiently at low redshift.} Mergers are most effective at increasing the star formation rate of the host galaxy at $z<1$, when the `ambient' level of star formation due to secular processes is low. 
\\
\item \textit{Both major and minor mergers enhance star formation, on average, by similar amounts at any given redshift.} e.g. minor mergers enhance SFRs, on average, by a factor of 1.69 at $z\sim2$, while the corresponding value for major mergers is 1.75. At $z\sim3.3$, minor mergers enhance SFRs, on average, by a factor of 1.68 while major mergers enhance SFRs by a factor of 1.69.
\\
\item \textit{Merger-driven enhancement of star formation decreases with increasing redshift.} While the merger rate increases with redshift, the SFR enhancement due to mergers decreases with redshift. This means that, while the fraction of star formation hosted \textit{in} merging systems increases with look-back time (due to the increasing merger rate), the fraction of star formation \textit{directly due} to mergers increases at a much slower rate.  
\\
\item \textit{Episodes of enhanced star formation typically occur prior to coalescence.} On average, 65 per cent of the enhanced star formation in a merger episode occurs prior to coalescence. Star formation in the post-merger remnant returns to levels found in the non-merging population on short timescales of around 1~Gyr. 
\\
\item \textit{Only $\fall$ per cent of the stellar mass growth since $z\sim6$ is directly attributable to galaxy mergers.} Major and minor mergers together account for just $\fall$ per cent of the stellar mass formed since $z=6$. Only $\sim\fmajor$ per cent of today's stellar mass is directly due to major mergers, while $\sim\fminor$ per cent is due to minor mergers. While individual minor mergers are less efficient enhancers of star formation, the minor merger rate outstrips the major merger rate at all redshifts, leading to a greater minor merger contribution over cosmic time. Thus, smooth accretion, not merging, is the dominant driver of stellar mass growth over the lifetime of the Universe.  
\end{itemize}

\section*{Acknowledgements}
We thank the anonymous referee for constructive comments that improved this paper. GM acknowledges support from the Science and Technology Facilities Council [ST/N504105/1]. SK acknowledges a Senior Research Fellowship from Worcester College Oxford. JD acknowledges funding support from Adrian Beecroft, the Oxford Martin School and the STFC. CL is supported by a Beecroft Fellowship. This research has used the DiRAC facility, jointly funded by the STFC and the Large Facilities Capital Fund of BIS, and has been partially supported by grant Spin(e) ANR-13-BS05-0005 of the French ANR. This work was granted access to the HPC resources of CINES under the allocations 2013047012, 2014047012 and 2015047012 made by GENCI and is part of the Horizon-UK project. James Geach, Donna Rodgers-Lee and Daniel Smith are thanked for useful comments.




\bibliographystyle{mnras}
\bibliography{references} 







\bsp	
\label{lastpage}
\end{document}